\documentclass[prl,twocolumn,amsmath,nobibnotes]{revtex4-1}

\usepackage[utf8]{inputenc}
\usepackage{graphicx, floatrow}
\usepackage[caption=false]{subfig}
\usepackage{amssymb, amsfonts, amsmath, dsfont}
\usepackage{comment}
\usepackage{multirow}
\usepackage{makecell}
\usepackage[T1]{fontenc}
\usepackage{floatrow}
\usepackage[table]{xcolor}
\usepackage{letltxmacro}
\usepackage[section]{placeins}

\usepackage[]{algorithm2e}
\usepackage{hyperref}

\DeclareRobustCommand{\Eq}[1]{Equation~\ref{eq:#1}}

\LetLtxMacro{\originaleqref}{\eqref}
\renewcommand{\eqref}[1]{\originaleqref{eq:#1}}

\def \>{\rangle} 
\def \<{\langle}

\def\be{\begin{equation}} 
\def\ee{\end{equation}} 
\newcommand \bea {\begin{eqnarray}} 
\newcommand \eea {\end{eqnarray}} 
 
\newcommand{\nn} {\nonumber}

\makeatletter
\newcommand*{\defeq}{\mathrel{\vcenter{\baselineskip0.5ex \lineskiplimit0pt
                     \hbox{\scriptsize.}\hbox{\scriptsize.}}}%
                     =}
\makeatother

\begin{document}

\title{A Rule of Thumb for the Power Gain due to Covariate Adjustment in Randomized Controlled Trials with Continuous Outcomes}

\author{Charles K. Fisher}
\email{drckf@unlearn.ai}
\affiliation{Unlearn.AI, Inc., 75 Hawthorne, San Francisco, CA 94105}

\date{\today}

\begin{abstract} 
Randomized Controlled Trials (RCTs) often adjust for baseline covariates in order to increase power. This technical note provides a short derivation of a simple rule of thumb for approximating the ratio of the power of an adjusted analysis to that of an unadjusted analysis. Specifically, if the unadjusted analysis is powered to approximately 80\%, then the ratio of the power of the adjusted analysis to the power of the unadjusted analysis is approximately $1 + \frac{1}{2} R^2$, where $R$ is the correlation between the baseline covariate and the outcome.
\end{abstract} 

\maketitle

Reference \cite{emaprocova} introduced a simple rule of thumb for approximating the relative power of a covariate adjusted analysis compared to an unadjusted analysis in an RCT with continuous outcomes. Specifically, if an adjusted analysis is performed using a baseline covariate that has a correlation $R$ with the outcome, then the ratio of the power in the adjusted analysis to the power in an unadjusted analysis is approximately $1 + \frac{1}{2} R^2$. This technical note provides a simple derivation of this formula, highlighting the regime in which it's a good approximation.

Consider an ANCOVA analysis of a 1:1 randomized trial with a constant treatment effect $\tau$ estimated while adjusting for a covariate with correlation $R$ with the outcome. The power to detect $\tau$ is
\begin{equation}
p = \Phi \left( \Phi^{-1} \left(\frac{\alpha}{2} \right) - \frac{\tau}{\nu} \right) 
+ 
\Phi \left(\Phi^{-1} \left(\frac{\alpha}{2}\right) + \frac{\tau}{\nu} \right) \nn
\end{equation}
in which 
\begin{equation}
\nu^2 = \frac{4 \sigma^2}{N} ( 1 - R^2 ) \, \nn
\end{equation}
is the asymptotic variance of the regression estimate for the treatment effect. If $\tau / \nu > 0$ then the second term in the formula for power dominates and
\begin{equation}
p \approx \Phi \left(\Phi^{-1} \left(\frac{\alpha}{2}\right) + \frac{\tau}{\nu} \right) \, .
\label{eq:power}
\end{equation}
Here, I've assumed the treatment effect is positive but the result is the same if the treatment effect is negative.

To fix a reference point, I consider the case in which the unadjusted analysis has power $\tilde{p}$ to detect $\tau$. From this, one can derive the sample size as 
\begin{align}
\sqrt{N} \approx \frac{\sqrt{4 \sigma^2}}{\tau} \left(\Phi^{-1} \left(\tilde{p}\right) - \Phi^{-1}\left(\frac{\alpha}{2}\right) \right) \, \nn
\end{align}
by setting $R=0$ in \Eq{power} and solving for $\sqrt{N}$. Plugging the sample size in to the formula for power yields
\begin{equation}
p \left(R^2 \right) \approx \Phi \left( \Phi^{-1} \left(\frac{\alpha}{2}\right) + \frac{\left(\Phi^{-1} \left(\tilde{p}\right) - \Phi^{-1}\left(\frac{\alpha}{2}\right)\right)}{\sqrt{ 1 - R^2 }} \right) \, . \nn
\end{equation}
The formula above is a good approximation under the standard assumptions for ANCOVA and the assumption of a constant treatment effect, but the various normal cumulative distribution functions make it a bit hard to understand.  

A simple rule of thumb can be derived from a series expansion about $R=0$. To simplify notation, I will define $a \defeq \Phi^{-1} \left(\frac{\alpha}{2}\right)$ and $b \defeq \Phi^{-1}\left(\tilde{p}\right) - \Phi^{-1}\left(\frac{\alpha}{2}\right)$ and rewrite the approximate formula for power as
\begin{equation}
p\left(R^2\right) \approx \Phi\left(a + \frac{b}{\sqrt{1 - R^2 }}\right) \, . \nn
\end{equation}
The ratio of the power of the adjusted analysis to the power of the unadjusted analysis is
\begin{equation}
\frac{p\left(R^2\right)}{\tilde{p}} \approx \frac{\Phi\left(a + \frac{b}{\sqrt{1 - R^2}}\right)}{\Phi\left(a + b\right)} \, . \nn
\end{equation}
To second order, the expression for the ratio of powers is
\begin{equation}
\frac{p\left(R^2\right)}{\tilde{p}} \approx 
\frac{bR^2e^{-\frac{1}{2}(a+b)^2}}{\sqrt{2\pi}\text{erfc}\left(\frac{-a-b}{\sqrt{2}}\right)}+\frac{2-\text{erfc}\left(\frac{a+b}{\sqrt{2}}\right)}{\text{erfc}\left(\frac{-a-b}{\sqrt{2}}\right)} + O(R^4)
\nn
\end{equation}
where $\text{erfc}$ is the complementary error function. Since clinical trials for pharmaceuticals often use $\alpha=0.05$ and unadjusted power of $\tilde{p} = 0.80$, I'll plug those values into the formula for the power ratio to yield the desired result
\begin{equation}
\frac{p \left(R^2\right)}{\tilde{p}} \approx 1 + \frac{1}{2} R^2 \, .
\end{equation}

In conclusion, in an RCT with continuous outcomes, the ratio of the power of an adjusted analysis to the power of an unadjusted analysis is approximately $1 + \frac{1}{2} R^2$, where $R$ is the correlation between the outcome and the baseline covariate. This simple rule of thumb is accurate when the treatment effect is constant, the correlation is relatively small, and when the unadjusted analysis is powered to approximately 80\%.

\bibliographystyle{plainnat}
\bibliography{refs}

\newpage
\appendix

\end{document}